\documentclass[journal=ancac3,manuscript=article]{achemso}

\usepackage{chemformula,dingbat} 
\usepackage[T1]{fontenc} 
\usepackage{soul}

\author{Athulya K. Muraleedharan}
\affiliation[LuMIn]
{Université Paris-Saclay, ENS Paris-Saclay, CNRS, CentraleSupélec, LuMIn, 91190 Gif-sur-Yvette, France}

\author{Jingye Zou}
\author{Maxime Vallet}
\author{Abdelali Zaki}
\author{Christine Bogicevic}
\affiliation[SPMS]
{Université Paris-Saclay, CentraleSupélec, CNRS, Laboratoire SPMS, 91190 Gif-sur-Yvette, France}

\author{Charles Paillard}
\affiliation[SPMS]
{Université Paris-Saclay, CentraleSupélec, CNRS, Laboratoire SPMS, 91190 Gif-sur-Yvette, France}
\alsoaffiliation[UA]
{Smart Ferroic Materials, Institute for Nanoscience \& Engineering and Department of Physics, University of Arkansas, Fayetteville 72701 Arkansas, USA}

\author{Karen Perronet}
\author{Fran\c cois Treussart}
\affiliation[LuMIn]
{Université Paris-Saclay, ENS Paris-Saclay, CNRS, CentraleSupélec, LuMIn, 91190 Gif-sur-Yvette, France}
\email{francois.treussart@ens-paris-saclay.fr}

\title[Electric field sensing with individual ferroelectric nanocrystal]
  {Sub-millisecond electric field sensing with an individual
rare-earth doped ferroelectric nanocrystal}


\begin{document}


\begin{tocentry}
\includegraphics{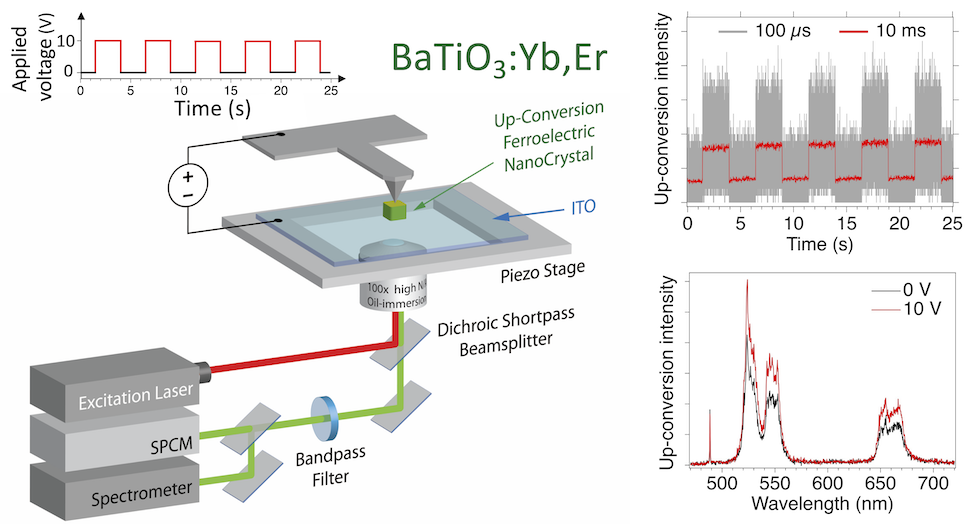}
Left: Electrical and optical scanning probe and fluorescence microscopy setup, to modulate and detect the up-conversion from single barium titanate nanocrystals doped with ytterbium and erbium ions (BaTiO$_3$:Yb,Er). Right top: intensity of up-conversion (UC) under a voltage alternating between 0 and 10 V applied to a single nanocrystal, demonstrating the ability to sense such variations with UC at 100~µs temporal resolution. Right bottom: Spectral variations associated to applied voltage of 10~V.

\end{tocentry}

\begin{abstract}
Understanding the dynamics of electrical signals within neuronal assemblies is crucial to unraveling complex brain function. Despite recent advances in employing optically active nanostructures in transmembrane potential sensing, there remains room for improvement in terms of response time and sensitivity. Here, we report the development of such a nanosensor capable of detecting electric fields with a submillisecond response time at the single particle level. We achieve this by using ferroelectric nanocrystals doped with rare earth ions producing upconversion (UC). When such a nanocrystal experiences a variation of surrounding electric potential, its surface charge density changes, inducing electric polarization modifications that vary, via converse piezoelectric effect, the crystal field around the ions. The latter variation is finally converted into UC spectral changes, enabling optical detection of electric potential. To develop such a sensor, we synthesized erbium and ytterbium-doped barium titanate crystals of size $\approx160$~nm. We observed distinct changes in the UC spectrum when individual nanocrystals were subjected to an external field via a conductive AFM tip, with a response time of 100~$\mu$s. Furthermore, our sensor exhibits a remarkable sensitivity of 4.8~kV/cm/$\sqrt{\rm Hz}$, enabling time-resolved detection of fast changing electric field of amplitude comparable to that generated during a neuron action potential.
\end{abstract}

\section{Introduction}
The generation and propagation of electrical signals in neurons and across neuronal assemblies, which occur at the nanometer and millisecond scales, remain poorly understood, despite the development of several theoretical models. For instance, the theory of core conductor cable and the Goldman-Hodgkin-Huxley-Katz models of ion motion and voltage propagation have provided theoretical foundations for neuronal physiology~\cite{Goldman.1943,Hodgkin.1952}. These theories account for the action potential, the few millisecond lasting voltage variation of 100~mV across the $\approx 5$~nm thick membrane (corresponding to 200~kV/cm peak voltage), which supports neuron-to-neuron communication. However, these models neglect electrodiffusion, that is, concentration changes associated with ionic currents. This assumption is true at millimeter scale like the squid giant axon but fails to describe accurately the processes taking place at the sub-micrometer size structure of synapses~\cite{Savtchenko.2017,Lagache.2019}. 
The Poisson-Nernst-Planck theory reveals that the geometry of synaptic compartments influences the distribution of local electric field generated during synaptic transmission, which subsequently changes the concentration of charged neurotransmitters undergoing electrodiffusion~\cite{Sylantyev.2008} and therefore has a functional impact. As the shape of the synapse changes during the execution of cognitive tasks~\cite{Magee.2020}, being able to measure the electric field on this scale would not only be an observable of such key plasticity but could provide an experimental test of the theory. There is thus a clear need to develop non-toxic, local methods to sense electric fields in biological media in order to refine current models of neuronal activity.

The detection of electric fields at the nanometer scale has initially been addressed using ultrahigh resolution nanopipette electrophysiology~\cite{Jayant.2017}. Although very powerful, this technique can only study a single synapse at a time. On the contrary, voltage indicators (VI) were designed to stain the entire neuronal membrane and allow fluorescence recording of transmembrane voltage dynamics~\cite{Peterka.2011} throughout the cell with optical diffraction resolution. Enormous progress has been made in the development of brighter and faster VI, and in particular thanks to the development of hybrid chemigenetic systems that allow to combine optimally chemical compounds with genetically encoded proteins~\cite{Knöpfel.2019}. However, subdiffraction imaging of a large number of synapses at high temporal resolution has not yet been reported. The development of a technology offering the necessary high spatiotemporal resolution (typically 1~ms, 200~nm) to reliably interrogate electrophysiological dynamics at multiple neuronal nanodomains would provide invaluable insights into open questions in neuronal electrophysiology.

Optically active nanostructures that transduce electric field changes into optical properties modifications constitute possible candidates, with photon-shotnoise ultimately limiting the sensitivity. 
In this domain, quantum dots (QD) are particularly attractive as their photoluminescence (PL) is redshifted by the quantum confined Stark effect, and in addition, with a core size of a few nanometers, they can fit within the neuron membrane lipid double layer where the field is maximal. Based on a QD-doped polymer layer, Rowland \textit{et al.} developed a device that can detect an electric field as small as 10~kV/cm within 1~ms time scale~\cite{Rowland.2015}, however obtained by involving about one billion of QD (estimated from QD concentration, sensing layer thickness and excitation laser spot size).
Another strategy of electric field optical sensing harnessed the coupling of an electroactive polymer to a metallic nanoantenna that transduces an external electric field into a detectable change in localized surface plasmon resonances~\cite{Habib2019}. Compared to sensing based on QD, this approach benefits from a four orders of magnitude larger interaction cross-section with the incident light, expected to translate into a higher sensitivity. Indeed with such a device, Habib \textit{et al.} were able to sense a stimulated electroactivity of cultured cardiomyocytes, with submillisecond time resolution~\cite{Habib2019} but to achieve such performances, the authors had again to integrate the light scattered by about a million nanoantenna, hence reducing the spatial resolution to about 1~mm.

However, to maintain diffraction-limited imaging, an electric field must be sensed by a single or very few nanoparticles only. There are a few examples of such systems that are based on electric-field modulation of charge or Förster resonant energy transfer (FRET) between the nanoparticle and another entity. 
For example, Nag \textit{et al.} used a QD-fullerene (C$_{60}$) conjugate, with the C$_{60}$ embedded in the cell membrane~\cite{Nag.2017}. At resting membrane potential the photoexcited QDs relax to the ground state partially by electron transfer to the C$_{60}$. This transfer increases upon cell membrane depolarization, leading to a PL intensity relative quenching $\Delta I/I_0$ of a few \% that provides a transmembrane potential variation readout both in cultured cell and in the cortex of live mouse. While chemical or electrical stimulation leads to QD PL modulation, the fastest response time was about 100~ms, two order of magnitude longer than the time response necessary to reproduce an action potential (AP) with fidelity.
Another conjugate introduced by Liu \textit{et al.} consists of an up-converting nanoparticle (UCNP) to which voltage sensing dyes were coupled, that serve on one hand to anchor the UCNP into cell membrane and are also FRET acceptors~\cite{Liu.20202e}. The authors of this study report a relative change $\Delta I/I_0$ in UCNP luminescence intensity in the 0.4-6.1\% range in cultured cells, depending on the stimulation conditions, and a time response of about 50~ms, still much longer than the duration of the AP.
Hence, the spatial ($\approx 200$~nm) and time (submillisecond) resolution required for membrane potential sensing remains to be addressed.

Here, we report the development of $\approx 160$~nm-sized standalone optically active nano-crystals (NCs) capable of sensing, at the level of a single particle, a field of $\approx 152$~kV/cm slightly smaller than the peak amplitude field of an action potential, with a millisecond response time.
We based our sensing principle on a specific UCNP made of a ferroelectric nanocrystal (NC) host doped with rare earth ions. When the NC is exposed to an external electric field, its lattice undergoes deformations due to a converse piezoelectric effect, which leads to spectral changes of the rare-earth ion up-conversion emission.
The effectiveness of this principle was established by Hao \textit{et al.} who demonstrated, by a varying electric field of 125~kV/cm amplitude, the modulation of the upconverted (UC) light from an ytterbium (Yb$^{3+}$) and erbium (Er$^{3+}$)-doped epitaxial barium titanate (BTO) thin film~\cite{Hao2011}, at room temperature (in BTO ferroelectric phase). The authors demonstrated an UC modulation at a frequency of 0.1~Hz, far below the maximum frequency of $\approx 10$~kHz limited by the UC process response time of $\approx 100~\mu$s in erbium-doped BTO~\cite{Mahata2015}. Thus, here we develop BTO:Yb,Er individual nanocrystals of size $\approx 160$~nm, which we expose to a fast varying electric field with the conductive tip of an atomic force microscope (AFM), coupled to a fluorescence microscope. We demonstrate that 47\% of the nanocrystals exhibit a fast variation in UC intensity with a response time of $\approx 100~\mu$s, limited by the duration of the UC optical cycle. Furthermore, BTO nanocrystals are ideal to develop sensors that could be rapidly tested in biological contexts due to their pre-existing use in drug delivery or tissue engineering~\cite{Sood.2023}, but also for their intrinsic optical two-photon excitation response\cite{Genchi.2016} (resulting from the crystal second-order nonlinear response) in cultured cells~\cite{Hsieh.2009,Sugiyama.2018,Jordan.2020} and larvae of small organisms~\cite{Pantazis.2010,Culic-Viskota.2012}, without or with grafted functional groups~\cite{Culic-Viskota.2012,Sugiyama.2018,Jordan.2020}.

\section{Results and discussion}
\subsection{Synthesis and characterization of erbium and ytterbium-doped BTO nanocrystals}
Erbium and ytterbium-doped BTO nanocrystals (nanoBTO:Yb,Er) were synthesized by a combination of coprecipitation and hydrothermal methods. Ytterbium was used to sensitize erbium up-conversion.
By adapting the Ba/Ti stoichiometry, we aimed at incorporating Er$^{3+}$ and Yb$^{3+}$ ions in A and B sites of the ABO$_3$ barium titanate perovskite structure respectively, because the ytterbium ions have the closest radius to the one of titanium (B site) in the six coordinated state~\cite{Xue.1988}.
We targeted the stoichiometry $\rm{Ba_{0.985}Er_{0.01}Ti_{0.925}Yb_{0.1}O_3}$, ensuring charge compensation, as described in Materials and Methods. The choice of an erbium concentration of 1\% was made based on Zhang \textit{et al.}\cite{Zhang.2011v6}, who showed that higher concentrations enhance cross-relaxation, leading to lower UC.

\begin{figure}[!ht]
\centering
  \includegraphics[width=\textwidth]{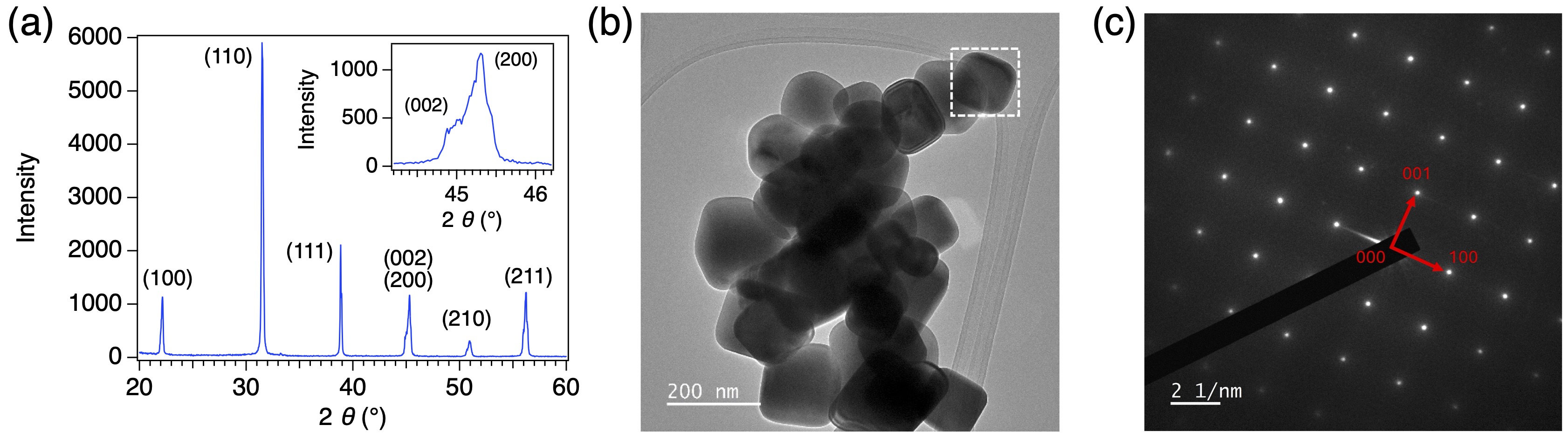}
  \caption{\textbf{Structural characterisation of erbium and ytterbium-doped barium titanate nanocrystals.} (a) XRD spectrum showing all detectable peaks of tetragonal BTO lattice. Inset graph: (002) plane peak splitting indicating tetragonality. (b) Bright-field TEM image of an aggregate of nanocrystals. Scale bar: 200~nm. (c)  Indexed SAED pattern corresponding to the particle squared with a dashed line in (b). We selected this particle because it was located on the edge of the aggregate,therefore allowing to position the TEM diaphragm to collect electrons diffracted by this nanocystal only. The high periodicity of this pattern indicates that the particle is a single crystal. Scale bar: 2~nm$^{-1}$}
  \label{fig:Structural_characterisation}
\end{figure}

The shape of the synthesized particles is mainly cubic with an average size of 158$\pm$27~nm, as deduced from scanning electron microscopy analysis (Supporting Information Figure~S1). Figure~\ref{fig:Structural_characterisation}a 
displays the powder X-ray diffractogram (XRD) at the room temperature of 293~K indexed to the crystallographic planes of the tetragonal phase of BTO according to the standard International Center for Diffraction Data (ICDD PDF card no.~05-0626)~\cite{Lee2012}. We did not observe any secondary phase structure, indicating the insertion of the doping ions in substitution of barium or titanium.
Rietveld refinement of this diffractogram (see Figure~S2) leads to $a=b=3.9980\pm 0.0001$~\AA\; and $c=4.0216\pm 0.0002$~\AA\; as the lattice parameters, hence $(c/a)_{\rm NC}=1.0059$ for the NC, a value lower than for the bulk BTO for which $(c/a)_{\rm bulk}=1.0101\pm 0.0002$~\cite{Megaw.1945}, indicating that NC lattice is slightly less tetragonal and more cubic than the bulk one, in agreement with other reports on undoped BTO NCs~\cite{Huang.2017ar,Lee2012}. The lower tetragonality of BTO NCs may result from a variety of factors, including the presence of lattice defects which are primarily hydroxide ions (OH$^-$) substituting to oxygen, coming from hydrothermal synthesis precursor materials, as well as the doped nature of our BTO NCs, which is known to often result in a smaller $c/a$ ratio~\cite{Hao2022}.

The powder diffraction analysis was confirmed by selected area electron diffraction (SAED) analysis using transmission electron microscopy (TEM). Figure~\ref{fig:Structural_characterisation}b displays a bright-field TEM image of a group of BTO nanocrystals, and Figure~\ref{fig:Structural_characterisation}c shows SAED pattern realized on a single BTO NC after one of its crystallographic axes has been aligned with the TEM electron beam. The diffractogram shows a regular pattern of bright spots, which are the result, in reciprocal space, of constructive interferences caused by electrons interacting with the periodic atomic structure of the crystal. The presence of a single rectangular pattern evidences that the particle is made of a single crystal. Furthermore, the dimensions of the rectangle are $1/a = 2.532\pm 0.010$~nm$^{-1}$ and $1/c = 2.516\pm 0.020$~nm$^{-1}$, leading to $a=0.395\pm 0.002$~nm and $c=0.397\pm 0.003$~nm, and a tetragonality value $c/a=1.006\pm 0.012$. These values are in good agreement with the ones derived from the XRD diffractogram (Fig.~\ref{fig:Structural_characterisation}a).

\begin{figure}[!ht]
\centering
  \includegraphics[width=\textwidth]{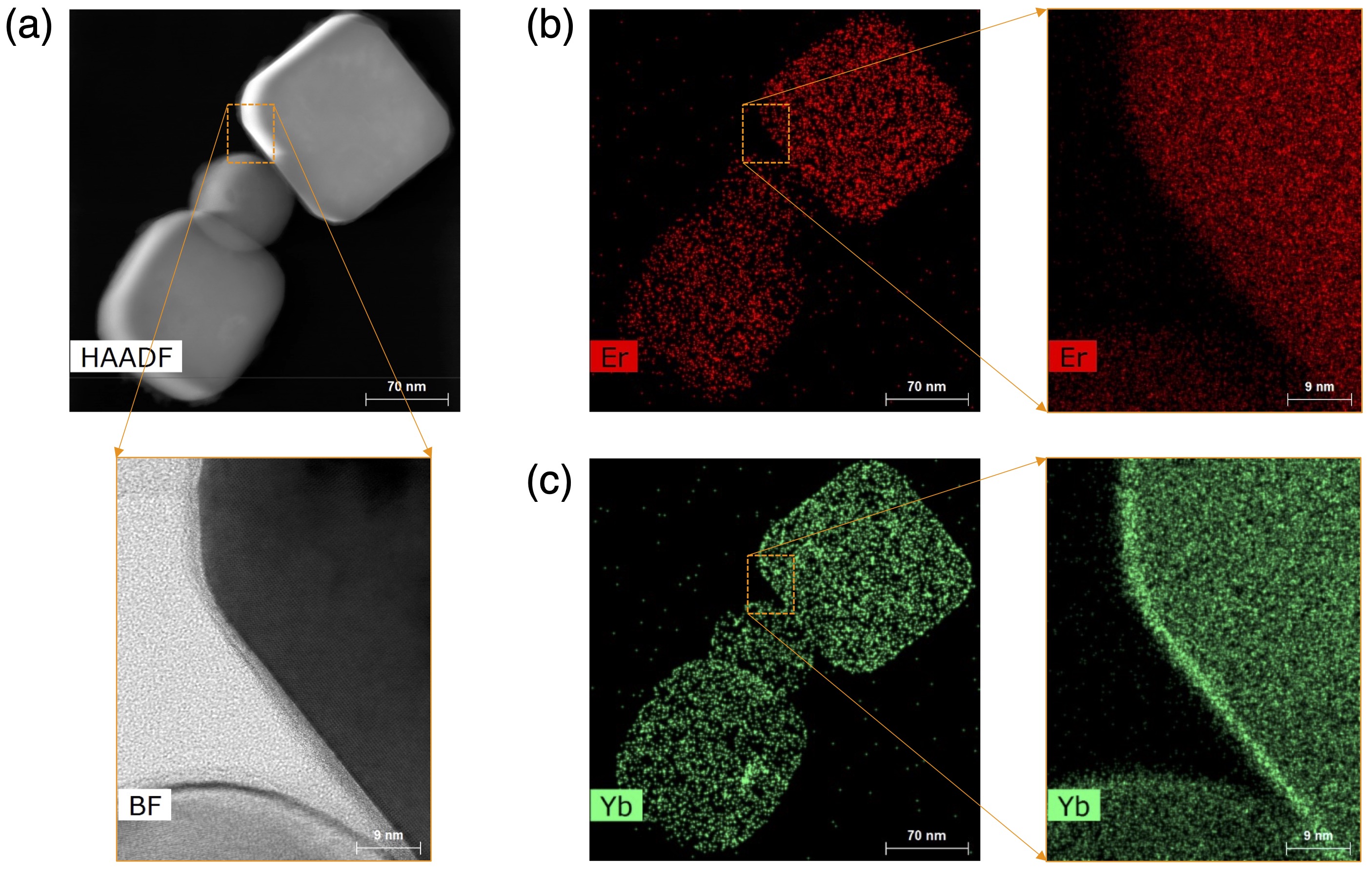}
  \caption{\textbf{Evaluation of ytterbium and erbium doping concentration by energy dispersive X-ray spectroscopy coupled to a scanning transmission electron microscope}. (a) HAADF STEM image of an aggregate of nanocrystals (scale bar: 70~nm) with a zoom, in bright field (BF), on corners of two particles (scale bar: 9~nm). (b) EDX image of the erbium distribution (scale bar: 70~nm) with a zoom (on the right, scale bar: 9~nm) on the same particle corners as in (a). (c) EDX image of the ytterbium distribution (scale bar: 70~nm), with a zoom (on the right, scale bar: 9~nm) on the same particle corners as in (a) and (b).}
  \label{fig:EDX}
\end{figure}
To measure the atomic fraction of erbium and ytterbium incorporated in the BTO nanocrystals, we conducted an elemental analysis at the individual particle level, using energy dispersive X-ray spectroscopy (EDX) coupled to the TEM (Materials and Methods). We aim to observe the characteristic EDX peaks of erbium at 6.947~keV and ytterbium at 7.414~keV in the particles. Figure~\ref{fig:EDX}a (top) shows the high-angle annular dark-field (HAADF) scanning TEM image of a group of three nanoBTO:Yb,Er. 
From the elemental analysis conducted on this group of NCs, we estimated (Materials and Methods) that erbium (Figure~\ref{fig:EDX}b) and ytterbium (Figure~\ref{fig:EDX}c) ions are incorporated at atomic concentrations of 1.7\% and 1.2\%, respectively. The erbium concentration is slightly higher than the targeted 1\%, while ytterbium effective concentration is significantly lower than the target of 10\%. In fact, the concentration of ytterbium is almost equal to that of erbium, likely to maintain charge neutrality by substituting both Ba\textsuperscript{2+} and Ti\textsuperscript{4+} ions equally by RE\textsuperscript{3+} ions, and thus avoiding the creation of charged defects such as oxygen vacancies.

One may wonder where Er\textsuperscript{3+} and Yb\textsuperscript{3+} ions insert themselves in the BTO nanocrystal. Previous works~\cite{Tsur2001,Vega2017} indicate that Yb\textsuperscript{3+} ions predominantly substitute Ti\textsuperscript{4+} ions, whereas Er$^{3+}$ can occupy both sites. For charge compensation reasons, it is thus very likely that Er\textsuperscript{3+} substitutes Ba\textsuperscript{2+}, but we cannot exclude that they also fill B sites~\cite{Buscaglia.2002}, which would lead to the appearance of oxygen vacancies. To further investigate effective substitution sites and the presence of such defects, other studies, such as UC lifetime~\cite{Mahata2015} and conductive measurements~\cite{Wang.2006cf} would be necessary but are beyond the scope of this article.

Increasing the effective content in Yb$^{3+}$ may lead to brighter NC. However, despite our efforts which consisted in varying multiple parameters during the synthesis (Ba/Ti ratio, known to influence Er$^{3+}$ site of incorporation~\cite{Buscaglia.2002}; erbium precursor compound; and ions concentrations between 0.25 and 2\% for Er$^{3+}$ and between 2.5\% and 13\% for Yb$^{3+}$), we were not able to introduce a larger proportion of ytterbium. The sample selected for the present study to test the electric field sensing is the brightest we achieved among the different synthesis we made. We noticed in high-resolution STEM of samples dimmer than the one selected for this article, the presence of polycrystalline ytterbium oxide nanoparticles surrounding the nanoBTO, which is consistent with a segregation of ytterbium outside the BTO host.

The characterization studies validated that the synthesized BTO nanocrystals are doped with ytterbium and erbium, and keep the tetragonal structure of undoped nanocrystals. Hence, nanoBTO:Yb,Er are very likely to maintain the ferroelectric and piezoelectric properties on which we based the electric field sensing mechanism. We then investigated the upconversion photoluminescence (PL) of individual nanoBTO:Yb,Er, and the modulation of its intensity by an external electric field.

\subsection{Up-conversion of single nanoBTO:Yb,Er and its modification by an electric potential}
\begin{figure}[!ht]
\centering
  \includegraphics[width=\textwidth]{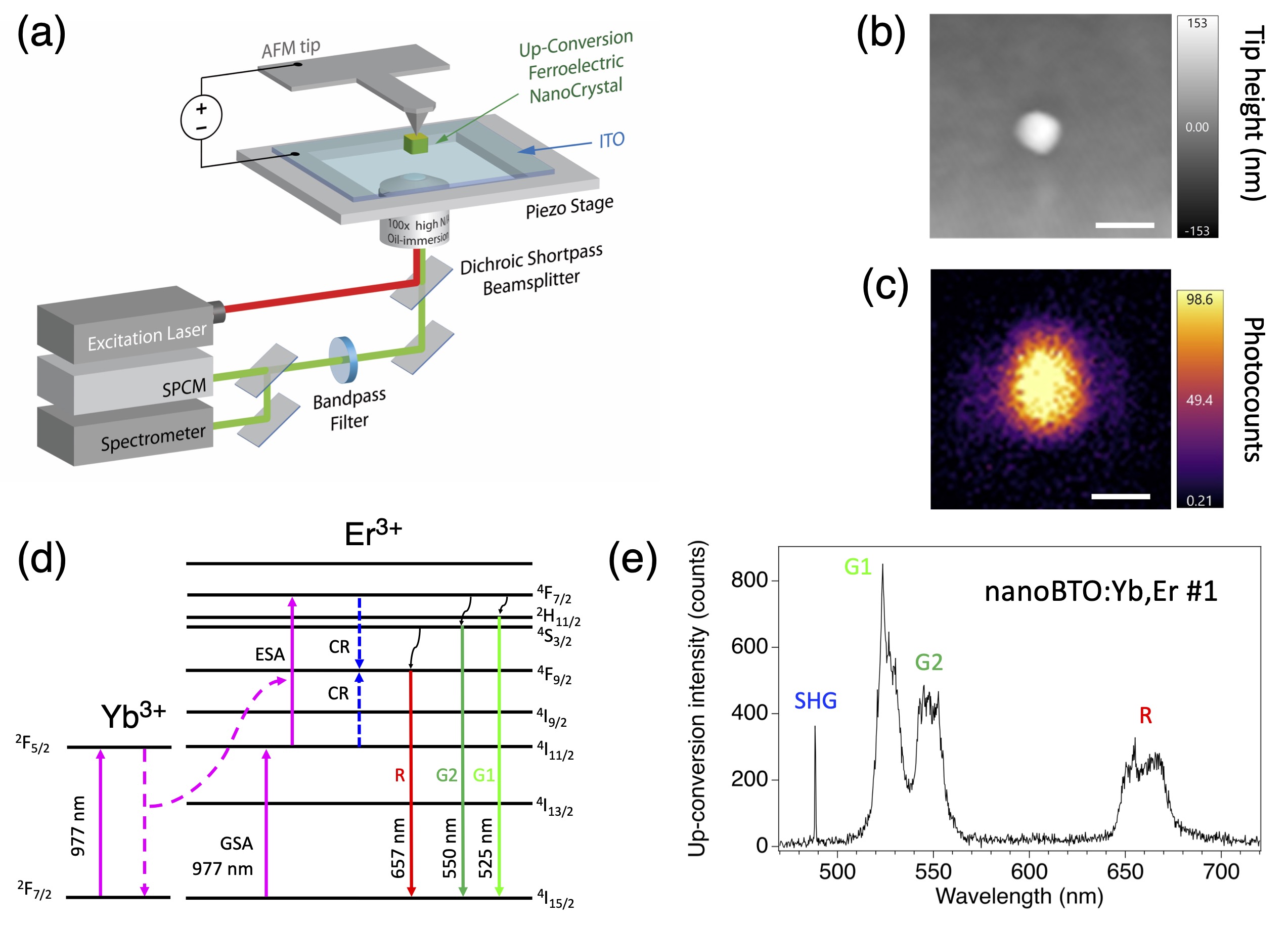}
  \caption{\textbf{Experimental setup to study single nanoBTO:Yb,Er luminescence changes under an applied electric field.} (a) Conductive AFM-fluorescence microscope correlated setup. The AFM head along with the AFM piezostage are positioned above the microscope objective of the fluorescence microscope. (b) Topography of single nanoBTO (nanoBTO:Yb,Er \#1) having a height of 153~nm (its lateral sizes are 180~nm$\times$180~nm from SEM image, not shown). (c) UC intensity scan of the same particle as in (b) acquired at 10~mW laser excitation power. Intensity scale: photocounts in 10~ms bin duration, yielding 9.86~kcounts/s maximum counting rate. Scale bars in (b)-(c): 500~nm. (d) Energy level diagram of ytterbium-erbium co-doped system showing Er$^{3+}$ upconversion transitions following direct laser excitation at 977~nm wavelength via ground state absorption(GSA) and excited state absorption (ESA), or GSA followed by resonant energy transfer from Yb$^{3+}$ in its excited state (dashed curved arrow) and subsequent ESA. The UC spectrum consists of three series of lines, G1, G2 and R corresponding to $^2$H$_{11/2} \rightarrow\, ^4$I$_{15/2}$,  $^4$S$_{3/2} \rightarrow\, ^4$I$_{15/2}$ and $^4$F$_{9/2} \rightarrow\, ^4$I$_{15/2}$ transitions, respectively. Blue dashed lines represent cross-relaxation (CR). (e) Up-conversion spectrum from the same single nanocrystal as in (b), under excitation laser power of 10~mW and a spectrometer Charged-Couple Device (CCD) array sensor integration time of 15~s. In addition to the up-conversion bands G1, G2 and R, we can also observe the SHG line at 488.5~nm wavelength.}
  \label{fig:UC_single_nanoBTO}
\end{figure}

To be able to apply a voltage directly on a single nanocrystal while recording its up-conversion PL, we used an AFM microscope in an electrical mode, coupled to a fluorescence microscope and we put the conductive AFM tip in contact with the top surface of a single nanoBTO:Yb,Er, with its bottom surface lying on a semi-transparent indium tin oxide (ITO) coated glass coverslip that is grounded (see Figure~\ref{fig:UC_single_nanoBTO}a, and Materials and Methods). To prevent the particle from being dragged by the tip while scanning and simultaneously ensure the electrical contact with ITO, we glued it to this substrate using poly(3,4-ethylenedioxythiophene) polystyrene sulfonate conductive polymer (Materials and Methods).
Once both microscopes are aligned relative to each other, topography (Figure~\ref{fig:UC_single_nanoBTO}b) and UC luminescence (Figure~\ref{fig:UC_single_nanoBTO}c) scans can be acquired simultaneously on the same particle.

The spectral analysis of single nanoBTO:Yb,Er luminescence under near infrared laser excitation (977~nm wavelength) revealed the three groups of lines in the visible range resulting from UC by erbium, as displayed in Figure~\ref{fig:UC_single_nanoBTO}d,e. These lines correspond to the transitions $^2$H$_{11/2} \rightarrow\, ^4$I$_{15/2}$ (wavelength around 525~nm with multiple peaks due to Stark splitting, labelled G1),  $^4$S$_{3/2} \rightarrow\, ^4$I$_{15/2}$ (around 550~nm, G2) and $^4$F$_{9/2} \rightarrow\, ^4$I$_{15/2}$ (around 657~nm, R).
The narrow peak in Figure~\ref{fig:UC_single_nanoBTO}e at 488.5~nm, half the excitation laser wavelength, is due to the second harmonic generation (SHG) by the NC, confirming the high crystallinity previously evidenced by XRD and SAED (Figure~\ref{fig:Structural_characterisation}a,c).

Note that 4f-4f electric dipole transitions of lanthanide ions are expected to be forbidden by quantum mechanics selection rules in a perfectly centrosymmetric crystalline environment, for which the odd orders of the crystal field vanish. The mere detection of erbium UC, even in absence of an external electric field, is the signature of lower local symmetry around erbium ions. In fact, the splitting of the $(hk\ell)$ Miller indexes (002) and (200) Bragg peaks near $2\theta=45^{\circ}$ in the inset of Fig.~\ref{fig:Structural_characterisation}a already indicates the presence of the polar (hence non-centrosymmetric) phase of BTO, enabling the 4f-4f electric dipole transitions of lanthanide ions. Note also that Zou \textit{et al.}~\cite{Zou.2023} reported erbium UC even in the high-temperature centrosymmetric phase of BTO, which indicates that the local environment felt by the RE ions is likely even less symmetric than the symmetry of the overall matrix, which is consistent with the eight-site model developed in BTO~\cite{Pirc2004} in which local symmetry breakings exist at the unit cell level but vanish on a macroscopic scale.

\begin{figure}[!ht]
\centering
  \includegraphics[width=\textwidth]{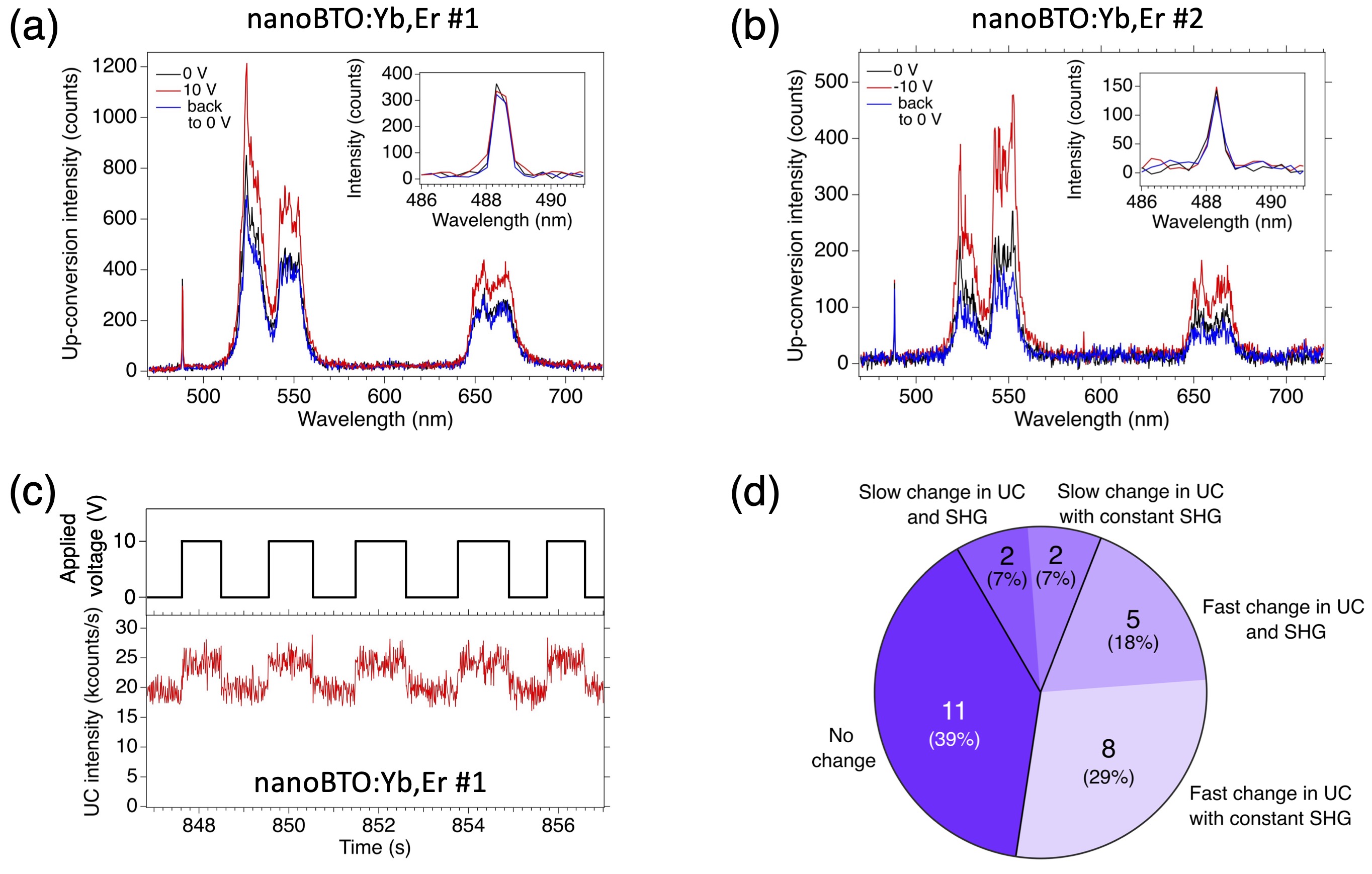}
  \caption{\textbf{Change of up-conversion spectrum of single nanoBTO:Yb,Er exposed to external electric field.} (a) UC spectra of nanoBTO:Yb,Er~\#1 (same NC as in Figure~\ref{fig:UC_single_nanoBTO}b-c,e), under an excitation laser power of 10~mW and spectrometer CCD integration time of 15~s, at voltage of 0~V (black line), +10~V (red), and back again at 0~V (blue). 
  (b) UC spectra of another particle, nanoBTO:Yb,Er \#2 (excitation laser power: 5~mW, spectrometer CCD integration time: 15~s) at voltage of 0~V (black line), -10~V (red), and back again at 0~V (blue). nanoBTO:Yb,Er \#2 height as measured by AFM (not shown) was 111~nm. 
  (c) Time trace of the total UC intensity of nanoBTO:Yb,Er \#1 (same as in (a)), while switching pseudo-periodically the applied voltage between 0 and +10~V. Bin time: 10~ms. 
  (d) Pie-chart distribution of the different behaviors observed for a total of 28 nanocrystals studied. The number of  particles per category is written in each section. NanoBTO:Yb,Er \#1 and \#2 are two of the eight particles displaying a fast change in UC (with constant SHG). The individual behaviors of these eight NC are summarized in Table~S1.}
  \label{fig:UC_change_static_E-field}
\end{figure}

We then investigated the changes of UC spectrum of a single nanoBTO:Yb,Er (nanoBTO:Yb,Er \#1) when we apply a static voltage of $\pm 10$~V across the particle. For the average particle size of 158~nm, this voltage corresponds to an electric field of $\approx 630$~kV/cm, more than twice smaller than the electric field breakdown of 1.5-3~MV/cm reported for thin BaTiO$_3$ in the literature~\cite{Wu.2020}. 
Figure~\ref{fig:UC_change_static_E-field}a shows UC spectral changes of nanoBTO:Yb,Er~\#1 under static voltage exposure of +10~V, manifested by an overall increase in the intensity of all peaks. This variation was reversible when the voltage was returned to 0~V, reverting to its initial value, and no change was detected with the opposite polarity of -10~V.
For another NC (nanoBTO:Yb,Er \#2), we also detected a reversible UC emission increase under an opposite applied voltage of -10~V (Figure~\ref{fig:UC_change_static_E-field}b), also with no change for the opposite polarity (of +10~V).
The intensities of the green and red UC emission bands both increase with the applied voltage, regardless of its polarity. This behavior differs from the one reported in BTO thin layer by Hao \textit{et al.}\cite{Hao2011}, where they observed an increase in the intensity of the green bands but none of the red one.
Hao \textit{et al.} proposed an interpretation based on the transition probability predicted by Judd–Ofelt theory, where the dominant contribution for the green emission transitions involves the intensity parameter $\Omega_2$ that increases when the symmetry of the site occupied by Er$^{3+}$ is reduced, which is what happens when the crystal is exposed to the external electric field. In contrary, the red emission transition does not involve this symmetry sensitive term and therefore should be almost insensitive to the electric field.
Our observation of a variation of intensity of the red emission transition under an external electric field indicates that in our sample, there is a more complex coupling of the red transition to the site symmetry changes. We hypothesize that the change of intensity of the red line could be due to an increase of population of the $^4$F$_{9/2}$ upper level of the red transition resulting from a larger cross-relaxation effect. The latter depends on the distance between ions, which indeed is modulated by the lattice deformation resulting from the converse piezoelectric effect, under the applied electric field.
Indeed we evidenced that in our nanoBTO:Yb,Er sample the red transition is governed by the cross-relaxation, because we observed that its intensity evolves linearly with the excitation laser intensity (see Figure~S3), consistently with two red photons emitted for two infrared excitation photons absorbed, with one of them experiencing a cross relaxation (see Figure~\ref{fig:UC_single_nanoBTO}d).

As for NC like nanoBTO:Yb,Er~\#1 and \#2 the intensities of both the green and red bands increase similarly under applied electric field, we considered the total UC intensity integrated over the whole spectrum. Figure~\ref{fig:UC_change_static_E-field}c displays this total UC intensity change under a voltage square modulation for nanoBTO:Yb,Er~\#1, if whole the UC light is sent to the single-photon counting module instead of the spectrometer. UC intensity modulation faithfully follows that of applied voltage.
We observed this overall behavior (fast reversible change of UC intensity) in 8 of the 28 NCs studied (Figure~\ref{fig:UC_change_static_E-field}d). 
As the SHG peak did not vary at all in these cases during the voltage application and removal process (see the inset of Fig.~\ref{fig:UC_change_static_E-field}b), we ruled out a change of focus of the microscope objective. Furthermore, the SHG peak stability also indicates that the applied electric field did not degrade the crystal lattice.
Among these eight particles displaying a fast change in UC with constant SHG signal, seven of them reacted to only one polarity (among which nanoBTO:Yb,Er \#1 and \#2, Figure~\ref{fig:UC_change_static_E-field}a,b) and a single one reacted to both polarities in an opposite manner. The responses of each of the eight nanocrystals to $\pm$10~V are summarized in Supporting Information Table~S1.  

To determine whether the change in UC intensity in nanoBTO:Yb,Er is indeed attributed to the ferroelectric properties of the crystal host as reported for BTO:Yb,Er thin film~\cite{Hao2011}, we conducted the following control experiments. First, we checked that the sole mechanical contact of the unbiased tip with the crystal does not induce a UC spectral change by piezoelectric effect (see Figure~S4). Then, we considered erbium-doped nanocrystals of yttrium oxide (Y$_2$O$_3$:Er, with 2\% doping). Y$_2$O$_3$ has a cubic crystallographic structure and therefore is devoid of piezoelectric and ferroelectric properties. Supporting Figure~S5 shows that $\pm10$~V applied to one Y$_2$O$_3$:Er$^{3+}$ nanocrystal (kind gift of Diana Serrano, synthesis done by Shuping Liu and detailed in Ref.\cite{Alqedra.2023}) did not induce any change of erbium up-conversion intensity. The absence of UC change under an applied voltage was confirmed for three other nanocrystals, evidencing that the ferroelectric BTO matrix is essential for the effect to happen. Finally, as one may wonder if the co-doping with Yb$^{3+}$ may play a key role in the transduction mechanism, we also tested NaYF$_4$ nanoparticles doped with 2\% erbium and 18\% ytterbium (kind gift of Thierry Gacoin, synthesis by Godefroy Leménager detailed in Ref.\cite{Leménager.2019}). We used nanocrystals in the $\beta$-NaYF$_4$ phase which structure belongs to the P63/m space group. The latter has polar properties, and therefore a NaYF$_4$ NC may also deform when exposed to an electric field. However, we did not observe any modulation of its UC under a variable applied voltage, as shown in Supporting Figure~S6. Hence, this additional experiment demonstrates the superior properties of ferroelectric BaTiO$_3$ NCs over another polar material such as NaYF$_4$:Yb,Er nanocrystals.

\begin{figure}[!ht]
\centering
  \includegraphics[width=\textwidth]{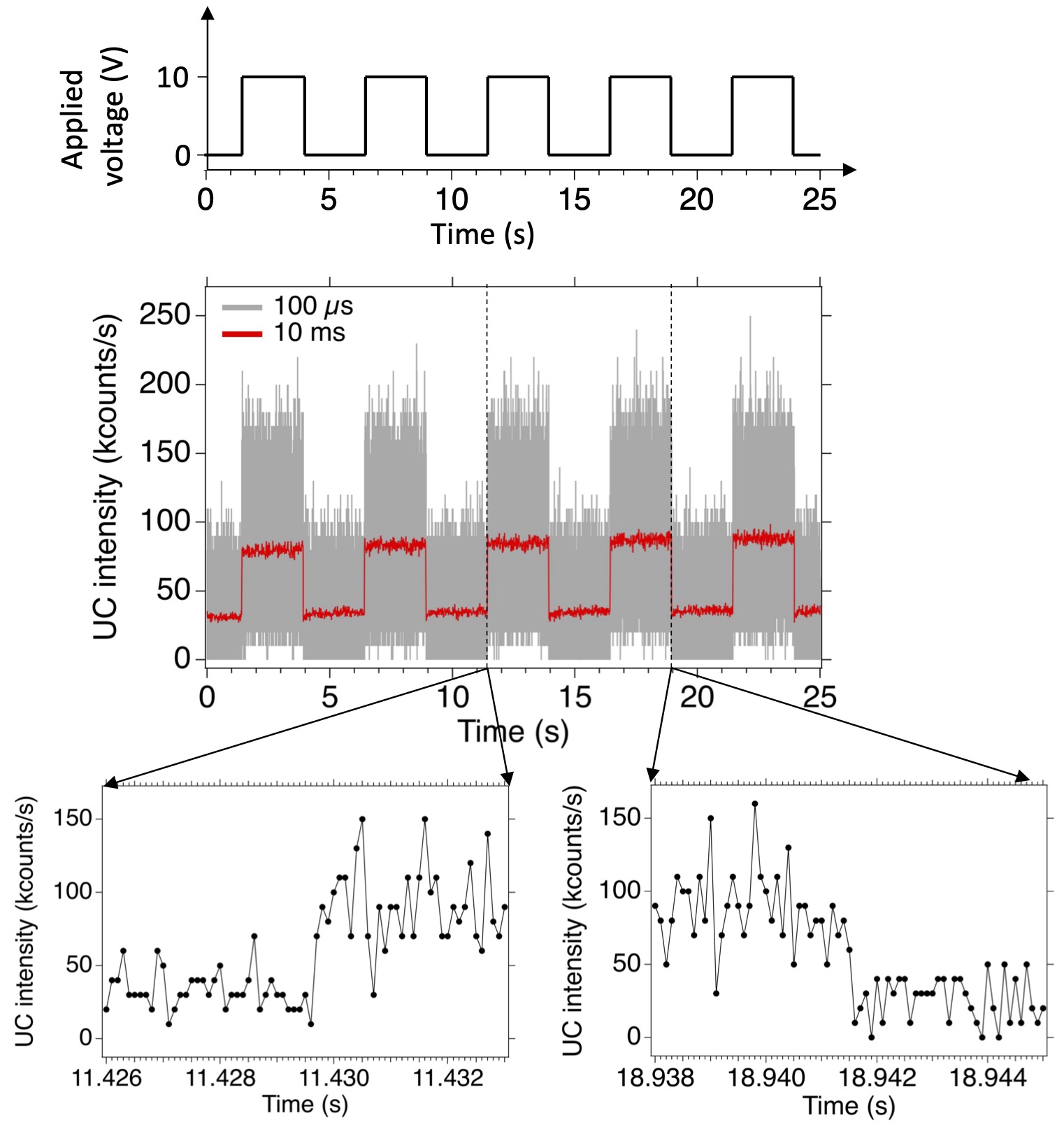}
  \caption{\textbf{Up-conversion intensity variation response time of a single nanoBTO:Yb,Er}. UC signal variation of a single 130~nm-sized doped-nanocrystal (nanoBTO:Yb,Er \#3) upon applied square voltage (+10~V) modulation (top trace), with 100~$\mu$s (grey) and 10~ms (red) bin duration (middle traces). Bottom: zooms on rising (left) and falling (right) edges showing UC intensity changes within a single time bin of 100~$\mu$s.}
  \label{fig:UC modulation}
\end{figure}

For each of these 8 particles with fast unipolar response, we then investigated the response time of its UC intensity variation when it was subjected to a modulated square electric potential (switching duration of only 1~$\mu$s, according to the bandwidth of the AFM electrical mode) with a frequency of 0.4~Hz and an applied voltage of 10~V. Figure~\ref{fig:UC modulation} displays, for one nanocrystal (nanoBTO:Yb,Er \#3), the total UC photocount rate time trace at bin durations of 100~$\mu$s and 10~ms. Based on the 10-ms time base, we observe a 2.7-fold increase of the photocount rate between 0 and 10~V. Moreover, zooming on the photocount rate trace at the rising and falling edges of voltage changes, we observe that it changes within a single time step of 100~$\mu s$, a value matching well the reported nanoBTO:Yb,Er up-conversion time response\cite{Mahata2015}. Furthermore, we estimated the signal-to-noise to be 1.6, calculated as the ratio of the average amplitude difference (between 0 and 10~V) to the standard deviation of UC intensity when 10~V is applied.  

The 20 other NCs studied exhibited different behaviors. First, 11 particles did not show any UC or SHG variation whatever the polarity was (Figure~S7a). In this case, bad physical contact between the tip and the NC top facet was ruled out because we engaged and withdrew the tip several times and obtained the same negative result. 
Two NC displayed slow (few seconds) changes in UC with no change in SHG (Figure~S7b); five displayed fast changes in UC and also in SHG (Figure~S7c), while two showed a slow change in UC and SHG (Figure~S7d). 
Note that we did not observe any correlation between the nanocrystals size and UC response: nanoBTO:Yb,Er\#1 to \#3, that displayed UC changes upon applied voltage variation, have measured height of 153, 111 and 130~nm respectively, while nanoBTO:Yb,Er of Figure~S7a, which did not manifest UC changes, has a height of 133~nm, similar to the one of reacting NC.

The diversity of behaviors could rather have its origin in at least two phenomena: electrical current leakage through the nanocrystal, preventing the electric field $\boldsymbol{E}$ to be applied, and/or the complex nanocrystal polarization texture that renders its crystal deformation under $\boldsymbol{E}$ dependent on the orientation of the field relative to the NC crystallographic axis $c$ as we recently evidenced in Ref.~\cite{Muraleedharan.2024}.
Current leakage has been reported in rare earth-doped ferroelectric thin films and was attributed to mobile oxygen vacancies appearing to ensure charge compensation. More precisely, it has been shown in (Ba,Sr)TiO$_3$ thin films, that when doping with a RE$^{3+}$ ion only on the A$^{2+}$ or on B$^{4+}$ site, the resulting charge imbalance generates oxygen vacancies responsible for a large increase in leakage~\cite{Wang.2006cf}. We hypothesize that the lack of electric field modulation of the UC may be due entirely or partly to a similar mechanism where Er$^{3+}$ and/or Yb$^{3+}$ mainly dope either the A$^{2+}$ or B$^{4+}$ site, but not both at the same time. Further experiments involving conductive-AFM and nanoscale analytical techniques (Electron Energy Loss Spectroscopy, EDX or Atom Probe Tomography~\cite{Gault.2021,Qu.2022}), outside of the scope of this article, are needed to test this hypothesis.

As for the other source of variable response related to the polarization texture, we have to take into account the orientation of $\boldsymbol{E}$ relative to the NC crystallographic $c$ axis.
Indeed, our geometrical configuration differs from the well defined one reported in Hao \textit{et al.}~\cite{Hao2011}, which involved BTO:Yb,Er doped thin film with $c$ oriented faces and $\boldsymbol{E}$ parallel to $c$. In this later case, $\boldsymbol{E}$ applied parallel to $c$ (being the direction $z$, indexed by 3) induces, by inverse piezoelectric effect, a strain $d_{33}E$, where $d_{33}$ is one of the piezoelectric coefficients of BTO. This strain is accompanied with ions relative displacement leading to variations of the crystal field around erbium and ytterbium ions, thereby changing the transition coefficients and subsequently UC intensity.

In the case of BTO NCs we recently showed by simulations and piezoresponse force microscopy that the polarization along the polar $c$ axis rotates by 90$^\circ$ over about 10~nm, to become in-plane within the $c$ facets, preventing a strong electrostatic depolarizing field. An electric field of amplitude $E$ applied to the $c$ facet by the AFM tip induces a shear strain displacement of amplitude $d_{15}E$, with $d_{15}=270$~pm/V being even larger than $d_{33}=191$~pm/V, according to Ref.\cite{Berlincourt.1958}. This shear strain is also expected to induce crystal field deformations leading to UC intensity changes. Furthermore, our simulations showed also that the lateral piezoresponse amplitude is one order of magnitude larger for $\boldsymbol{E}$ perpendicular to $c$ plane than to $a$ or $b$ planes~\cite{Muraleedharan.2024}. This dependence of the deformation amplitude to the electric field orientation relative to the $c$ axis, may contribute to the diversity of UC variation observed in the present work.

Polarization-resolved second-harmonic generation radiative pattern measurement is conventionally used to determine the crystalline axis orientation of a monocrystal, even nanometric size ones~\cite{Brasselet.2003}. However, this method does not work with BTO nanocrystals due to their complex polarization texture~\cite{Muraleedharan.2024}. Hence we are not able to determine the $c$ axis orientation of the individual NC probed for their electric-field sensing ability, which prevents us to check the influence of  $\boldsymbol{E}$ orientation relative to $c$ on the signal intensity.

\subsection{Sensitivity of the nanoBTO:Yb,Er electric field sensor}
With nanoBTO:Yb,Er \#3, of size 130~nm, we were able to detect a field amplitude $E_{\rm min}\equiv 10/(130\times 10^{-7})\approx 770$~kV/cm within $T=100~\mu$s with a signal-to-noise ratio of 1.6. The minimum field detectable during an optical cycle is then $E_{\rm min}=770/1.6\approx 480$~kV/cm, leading to a sensitivity $S_{\rm 1nanoBTO}\equiv E_{\rm min}\sqrt{T}=4.8$~kV/cm/$\sqrt{\rm Hz}$.
Considering this sensitivity, if we integrate over 1~ms (a third of the total duration of a neuron action potential), we should be able to detect a field $E=S_{\rm 1nanoBTO}/\sqrt{10^{-3}}\approx 152$~kV/cm. This value is slightly smaller than the peak-to-peak axonal membrane potential variation of $\approx 200$~kV/cm during an action potential, opening prospects in the detection of such an event.
Moreover, the sensitivity should improve by considering larger particles as the signal increases with more ions being exposed to the electric field.

We now compare the nanoBTO:Yb,Er sensor sensitivity to that of other single nanostructures envisioned to sense trans-membrane potential fast changes mentioned in the introduction. In particular, PL modulation of QDs attributed to electric field ionization was reported to detect a minimum field of $E_{\rm min}^N\approx 10$~kV/cm with 1~ms time resolution~\cite{Rowland.2015}, but it involved $\approx 10^9$ QDs. Hence, we extrapolate the single QD sensitivity of this system to be $S_{\rm 1QD}\approx 10\times \sqrt{10^9}\times\sqrt{10^{-3}}=1000$~kV/cm/$\sqrt{\rm Hz}$, which is 208 times less sensitive than a single nanoBTO:Yb,Er sensor. 
Similarly, the FRET couple formed by a voltage detecting dye and an up-conversion nanoparticle~\cite{Liu.20202e} is capable of reporting an electrophysiological signal by variation in relative intensity, but it also requires a large number of probes and has a slow response time of 50~ms, which prevents to resolve the temporal profile of an individual action potential. 
Finally, the other optical signal-based neuron activity sensing approach recently reported, harnessing electroactive polymer coupled to plasmonic nanoantenna~\cite{Habib2019}, yields both a very low detection threshold of 0.1~kV/cm and a fast (200~$\mu$s) response, but here also, it requires the accumulation of the signal from a large number (one million) of antenna reducing the spatial resolution down to 1~mm. Furthermore, the extrapolated single antenna sensitivity is $S_{\rm 1antenna}=0.1\times\sqrt{10^6}\times\sqrt{0.2\times 10^{-3}}\approx 1.4~$kV/cm/$\sqrt{\rm Hz}$. Although this extrapolated $S_{\rm 1antenna}$ is better than $S_{\rm 1nanoBTO}$, its experimental realization has not yet been demonstrated, contrary to that of a single nanoBTO:Yb,Er.

\section{Conclusions}
Fluorescent voltage indicators, including the advanced genetically encoded ones (GEVI), have become the gold standard in optical recording of electrophysiological activity of an assembly of neurons, as they provide the millisecond time scale resolution required for a faithful rendering of electrical potential variations~\cite{Knöpfel.2019}. However, to yield a sufficiently strong fluorescence, GEVI need to be over-expressed, which may impact neuron physiology. Moreover, like all dyes, the fluorescent protein(s) in the GEVI photobleach(es) over time. Finally, interrogating electrical response with VI or GEVI at synaptic nanodomains remains challenging. For these reasons, alternative nanosensors are still worth developing, as the rare earth ions-doped ferro/piezo-electric nanocrystals we introduced in this article, that are capable of transducing an applied electric field in changes in the up-conversion emission intensity.

We synthesized, by hydrothermal method, erbium and ytterbium-doped barium titanate NCs of average size 158~nm, with effective doping estimated to be 1.7\% and 1.2\% for Er$^{3+}$ and Yb$^{3+}$ respectively. Despite the doping, the nanoBTO retained its tetragonal structure to which ferro- and piezo-electric properties are associated at room temperature of 293~K. We observed that 61\% of the nanocrystals displayed a variation of up-conversion intensity upon application of +10~V, -10~V or rarely for both polarities. Moreover, about half (47\%) of the nanocrystals had a fast response  (characteristic duration <100~$\mu$s), compatible with a faithful sampling of action potential temporal profile. UC intensity changes are attributed to fast modifications of the crystal field around erbium ions, that lower the symmetry inducing larger intra-4f transitions probabilities. We suggested that the detectable UC change may be due either to some current leakage through the particle or/and to unfavorable crystallographic axis orientation relative to the applied electric field. 

Mobile oxygen vacancies are considered to be responsible for current leakage in perovskite oxide ferroelectrics ABO$_3$, strongly affecting their ferroelectric and piezoelectric properties. In rare-earth ions doped ABO$_3$ crystals, such vacancies appear to ensure charge neutrality in circumstances when RE$^{3+}$ substitute to A and/or B sites without compensating the charges. Conduction measurements on individual NC will help to state on the presence of such vacancies~\cite{Wang.2006cf}, and if it is confirmed, the synthesis will need to be further improved.

Regarding the influence of the NC orientation relative to the electric field on sensing, we established in a recent piezoresponse force microscopy study~\cite{Muraleedharan.2024}, that the PFM displacement amplitude is one order of magnitude smaller for NC with their $c$ axis perpendicular the applied electric field than for $c$ parallel to the field (\textit{i.e.}, perpendicular to the substrate).
It is desirable to make the nanosensor more isotropic and less sensitive to the electric field orientation. As the largest crystal deformations induced by the electric field take place at the NC surface, we envision that core-shell perovskite structures (like the one reported in Ref.~\cite{Fujii.2013}) may be grown by epitaxy in such a way that shell $c$ axis orientation is orthogonal to all facets. In such composite, the in-plane crystal deformation should be large and independent of the external field direction.

With the nanoBTO:Yb,Er sensor we developed, we demonstrated a single particle sensitivity to electric field sensing of 4.8~kV/cm/$\sqrt{\rm Hz}$, corresponding to the ability to distinguish a field of 152~kV/cm from the noise, with a millisecond temporal resolution. As the peak-to-peak field amplitude of 200~kV/cm (during the 1~ms duration of an action potential) remains only slightly larger than the sensor sensitivity, improving the latter is desirable. To do so we could try to enhance erbium emission. In particular, we could incorporate lithium ions in addition to erbium and ytterbium, as it has been shown that lithium (by either A site substitution or by interstitial insertion) can increase the UC intensity by further lowering crystal symmetry around Er$^{3+}$ ions~\cite{Chen.2011}. Furthermore, a very thin shell (thickness<5~nm) of silanized ytterbium complex could be used to enhance the absorption cross-section of the near-infrared excitation light by the nanocrystal.

The application of the nanoBTO:Yb,Er to neuron cell membrane potential variation sensing requires the sensor to be brought into contact with the membrane as the Debye layer electric field shielding takes place within a few nanometers. We envision that fulfilling this localization constrain is much more critical for sensing than controlling the nanocrystal crystalline axis orientation relative to the electric field.
Such membrane localization may be achieved by surface functionalization of the nanocrystals. Indeed, grafting of bioconjugates to BTO has been reported by several groups~\cite{Sugiyama.2018,Jordan.2020,Pantazis.2010}. Following similar strategies, our collaborators are currently functionalizing the nanocrystals with polyethylene glycol (PEG)-cholesterol conjugates, able to anchor the sensor in the cell membrane and at the same time, thanks to the PEG chains, prevent the adsorption of culture medium proteins that would lead to non-specific interactions.
	
Finally, the nanoBTO:Yb,Er electric field sensor may also find applications in other scientific areas where electric field need to be probed with nanometer and sub-millisecond resolutions at room temperature. In these domains, the nitrogen-vacancy center in a diamond nanopillar offers a record sensitivity of only 0.24~kV/cm/$\sqrt{\rm Hz}$, but its implementation is still complex~\cite{Huxter.2023}.

\section{Materials and Methods}
\subsection{Synthesis of erbium and ytterbium-doped BTO nanocrystals}
The Yb/Er-doped $\rm{BaTiO_3}$ nanocrystals were prepared via the hydrothermal method in a designed stoichiometric ratio of $\rm{Ba_{0.985}Er_{0.01}Ti_{0.925}Yb_{0.1}O_3}$. Note that the chosen composition allows charge compensation as the total positive charge is equal to $0.985\times(+2)+0.01\times(+3)+0.925\times(+4)+0.1\times(+3)=+6$, opposite to the total negative charges of $3\times(-2)=-6$. 
We used barium carbonate (BaCO$_3$) and titanium isopropoxide (Ti(O\textit{i}Pr)$_4$) as the starting materials. BaCO$_3$ was dissolved in diluted acetic acid and warmed on a hot plate of a magnetic stirrer for complete dissolution. Er(CH$_3$COO)$_3$ and Yb(NO$_3$)$_3$ were added subsequently. The solution was then cooled down to room temperature, Ti(O\textit{i}Pr)$_4$ was slowly added, and the mixture was agitated during 2~h. Subsequently, the solution was dried in the oven at 100$^{\circ}$C for 6~h and transferred to a Teflon-lined autoclave, accompanied by the NaOH. The autoclave was kept in an oven for hydrothermal reaction at 220$^{\circ}$C for 24~h and naturally cooled down to room temperature in air. The collected product was washed several times with distilled water and dried at 100$^{\circ}$C for 12~h.

\subsection{Characterization instruments and methods}
X-Ray diffraction spectrum was acquired using a Bruker D2 diffractometer using Cu K$\alpha$1 radiation (wavelength: 1.5418~\AA) and Cu K$\alpha$2 radiation (wavelength: 1.5444~\AA). XRD data were collected at room temperature between angles 20$^\circ$ and 60$^\circ$ by steps of 0.02$^\circ$. The EVA software was then used to determine the phase composition of the material. Transmission electron microscopy (TEM) analyses of Figure~\ref{fig:Structural_characterisation}b-c was done with a JEOL JEM-2100Plus operated at 200~kV. \st{in parallel and in scanning mode (TEM-STEM) Scanning TEM (STEM) acquisition} were conducted utilising a FEI Thermofisher Titan3 G2 80-300 microscope, operated at 300~kV and equipped with a Cs probe corrector, STEM detectors, an Ultrascan 1000 XP TEM camera and a SuperX detector for energy-dispersive X-ray (EDX) analyses. EDX quantifications have been performed by using the Cliff-Lorimer method~\cite{Cliff.1975}. As Ti-K$_\alpha$ (4.510~keV) and Ba-L$_\alpha$ (4.465~keV) rays have very close energies, an optimisation of $k$-factors has been performed using pure BTO.

\subsection{Electrical mode AFM coupled to a fluorescence microscope}
The AFM (Asylum MFP-3D, Oxford Instruments) was coupled to a homemade fluorescence microscope based on a commercial microscope stand (Eclipse TE300, Nikon). 
A conductive ASYELEC-01-R2 tip with titanium/iridium coating (spring constant: 2.8~N/m) was used to apply a voltage on the nanocrystal. In this AFM model, the tip is static and the sample is scanned laterally using a piezostage, whose position can be finely adjusted relative to the microscope stand so that the AFM tip coincides with the luminescence detection volume.
The upconversion from nanoBTO:Yb,Er was excited using a continuous-wave fibered Bragg grating laser diode operating at 977~nm wavelength (BL976-SAG300, Thorlabs). The laser beam is sent through an apochromatic $\times$100 magnification and 1.49 numerical aperture oil immersion microscope objective (MRD01991, Nikon) by a short-pass dichroic mirror (700dcsx, Chroma Technology) which reflects it, and transmits the UC signal. The latter is then sent to a single photon counting module detector (SPCM AQR-14, PerkinElmer) through a band-pass filter of transmission range 450-700 nm (FF01-715/SP, Semrock) to filter out residual excitation laser light, thanks to an optical density larger than 6 at 977~nm. The UC spectrum is acquired using a 140~mm focal length imaging spectrometer (MicroHR, Jobin-Yvon/Horiba) using a 600~lines/mm grating blazed at 500~nm. The detector attached to this spectrograph is a cooled (sensor temperature of -70$^\circ$C) front-illuminated open electrode CCD array (Symphony, Jobin-Yvon/Horiba). All the measurements were performed on samples maintained at room temperature.

\subsection{Immobilization of nanoBTO}
A 1\% weight poly (3,4-ethylenedioxythiophene) polystyrene sulfonate (PEDOT:PSS) in water suspension was spin coated on a 170~$\mu$m thick cover glass covered with a semitransparent layer of ITO (80~nm ITO layer thickness), forming a $\approx 50$~nm thick layer, on top of which the nanoBTO:Yb,Er aqueous suspension is deposited. Due to the good wettability of BTO by water~\cite{Li2014}, the concave meniscus on the edges of the NCs drags them down by surface tension, leading to a final position of the particle emerging from the polymer layer as we showed in another work~\cite{Muraleedharan.2024}.

\section*{Author contributions}
Conceptualization: CP, CB, KP and FT; Methodology: AKM, JZ, CB, MV, AZ, KP, and FT; Validation: AKM, KP and FT; Investigation: AKM, JZ, MV, KP and FT; Resources: CB and JZ; Data curation: AKM, MV, AZ, and FT; Writing – Original draft: AKM and FT; Writing – Review and Editing: CP, AKM, KP, and FT; Visualization: AKM, MV, AZ, KP, and FT; Supervision: FT; Project administration: FT; Funding acquisition: CP and FT.
All authors have read and agreed to the published version of the manuscript.


\begin{acknowledgement}
The authors thank Shuping Liu and Godefroy Leménager for the synthesis of non ferroelectric Y$_2$O$_3$:Er and of NaYF$_4$:Yb,Er nanocrystals, respectively, used for the control experiments, and Céline Fiorini-Debuisschert for her help in finding a solution to firmly attach the nanocrystals to the conductive coverglass substrate. We also thank Brahim Dkhil, Aleix G. Güell, Julien Boudon, Lionel Maurizi and Nadine Millot for fruitful discussions.
The AFM used in this work was purchased thanks to the FOSTER program of ENS Paris-Saclay.
This work has received financial support to B.D. and F.T. from the CNRS through the MITI interdisciplinary program and from the French National Research Agency (ANR, grant numbers ANR-21-CE09-0028 and ANR-21-CE09-0033).
\end{acknowledgement}

\begin{suppinfo}
Supporting information file contains Supporting Data Figures: Size and shape distributions of nanoBTO:Yb,Er; Rietvelt refinement analysis on nanoBTO:Yb,Er; Variation of intensity of the green and red up-conversion bands with the laser excitation power; Absence of UC spectrum modification on nanoBTO:Yb,Er due to the AFM tip mechanical force applied at the contact; Absence of UC spectrum modification under applied voltage in non-ferroelectric Y$_2$O$_3$:Er nanocrystals; Examples of other up-conversion signal variation behaviors with applied electric field, differing from the fast variation with constant SHG.
\end{suppinfo}
\providecommand{\latin}[1]{#1}
\makeatletter
\providecommand{\doi}
{\begingroup\let\do\@makeother\dospecials
	\catcode`\{=1 \catcode`\}=2 \doi@aux}
\providecommand{\doi@aux}[1]{\endgroup\texttt{#1}}
\makeatother
\providecommand*\mcitethebibliography{\thebibliography}
\csname @ifundefined\endcsname{endmcitethebibliography}
{\let\endmcitethebibliography\endthebibliography}{}

\end{document}


\renewcommand{\figurename}{Figure~S\kern -3.5pt}

\renewcommand{\tablename}{Table~S\kern -3pt}

\vfill
\begin{figure}[H]
	\centering
	\includegraphics[width=0.9\textwidth]{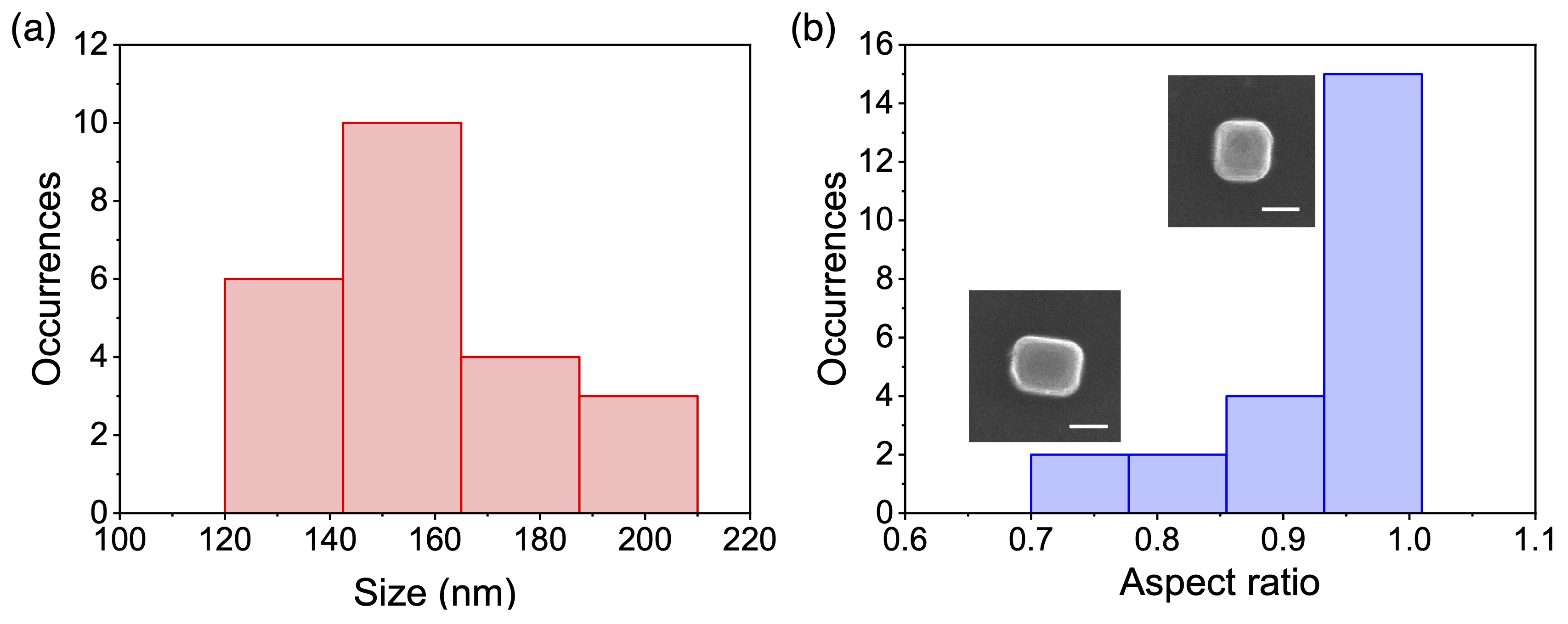}
	\caption{\textbf{Size and shape distributions of nanoBTO:Yb,Er}. (a) Particle size distribution as inferred from SEM images (size defined as the arithmetic mean of the side lengths of the rectangular shaped nanocrystals) with an average value of $158\pm 27$~nm. (b) Aspect ratio, revealing a majority of particles with a ratio of 1 (cubic shape). Inset: SEM images (30~kV acceleration voltage and 100000$\times$ magnification) of single NCs with aspect ratio 1 and 0.75; scale bars: 100~nm.}
	\label{SI_Fig:Size_shape_dist}
\end{figure}

\vfill
\begin{figure}[H]
	\centering
	\includegraphics[width=0.8\textwidth]{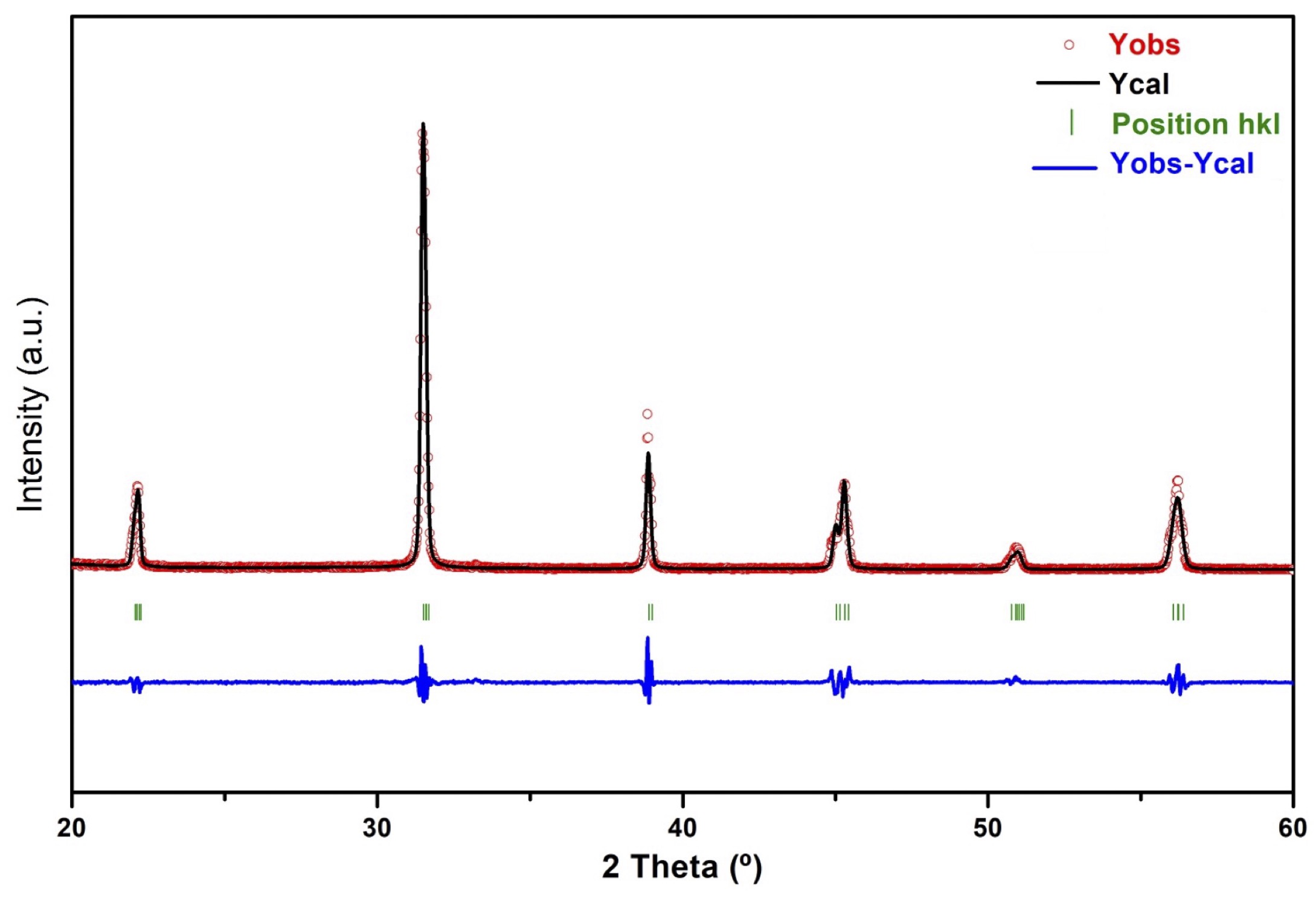}
	\caption{\textbf{Rietvelt refinement analysis on nanoBTO:Yb,Er}. Experimental diffractogram data (red open circles), modeled diffractogram (black line), small green vertical bars below the diffractogram indicate $hk\ell$ positions and the blue curve beneath represents the difference between the data and the model. The refinement was carried out using the WinPlotr/FullProf package~\cite{Rodriguez-Carvajal1993}. The peak shape was described by a pseudo-Voigt function, and the background level was modeled using a polynomial function.}
	\label{SI_Fig:Refinement}
\end{figure}

\begin{figure}[H]
	\centering
	\includegraphics[width=\textwidth]{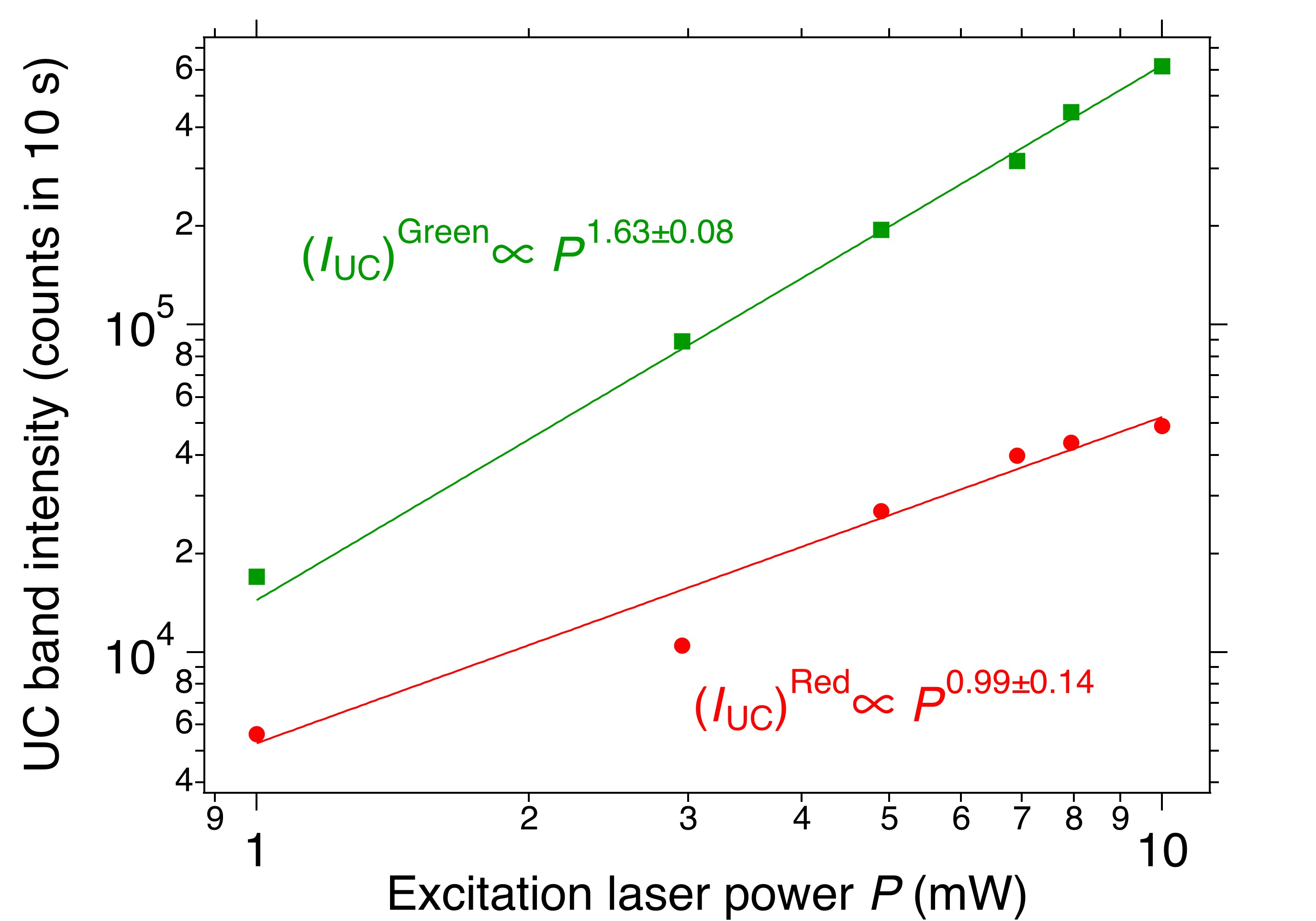}
	\caption{\textbf{Variation of intensity of the green and red up-conversion bands with the laser excitation power}. Total area of all the green transition bands (green squares, $(I_{\rm UC})^{\rm Green}$) and the red transition band (red circles, $(I_{\rm UC})^{\rm Red}$) as a function of the infrared laser (977~nm wavelength) excitation laser $P$. Lines are fits to a power law. $(I_{\rm UC})_{\rm red}$ has a linear dependence with $P$.}
	\label{SI_Fig:green_and_red_bands_vs_excitation_power}
\end{figure}

\begin{table}[H]
	\centering
	\begin{tabular}{l|cccccccc}
		&  1& 2& 3& 4& 5&  6&  7& 8\\\hline\hline
		UC$\nearrow$ with +10~V&  \checkmark&  &  \checkmark&  &  &  &  & \\\hline
		UC$\searrow$ with +10~V&  &  &  &  slow&  \checkmark&  slow&  \checkmark& \checkmark\\\hline
		UC$\nearrow$ with -10~V&  &  \checkmark&  &\checkmark&  slow&  \checkmark&  \checkmark& \\\hline
		UC$\searrow$ with -10~V&  &  &  &  &  &  &  & \\\hline
	\end{tabular}
	\caption{\textbf{Different responses to positive or negative applied voltage for the eight nanoBTO:Yb,Er (same numbering as in the main text) displaying fast up-conversion intensity variation}. The checkmark indicates observed fast variation (within one time bin duration of 100~$\mu$s), while ``slow'' means a response lasting up to a few seconds.}
	\label{tab:different_behavior_for_nanoBTO_with_fast_varying_UC}
\end{table}

\begin{figure}[H]
	\centering
	\includegraphics[width=0.7\textwidth]{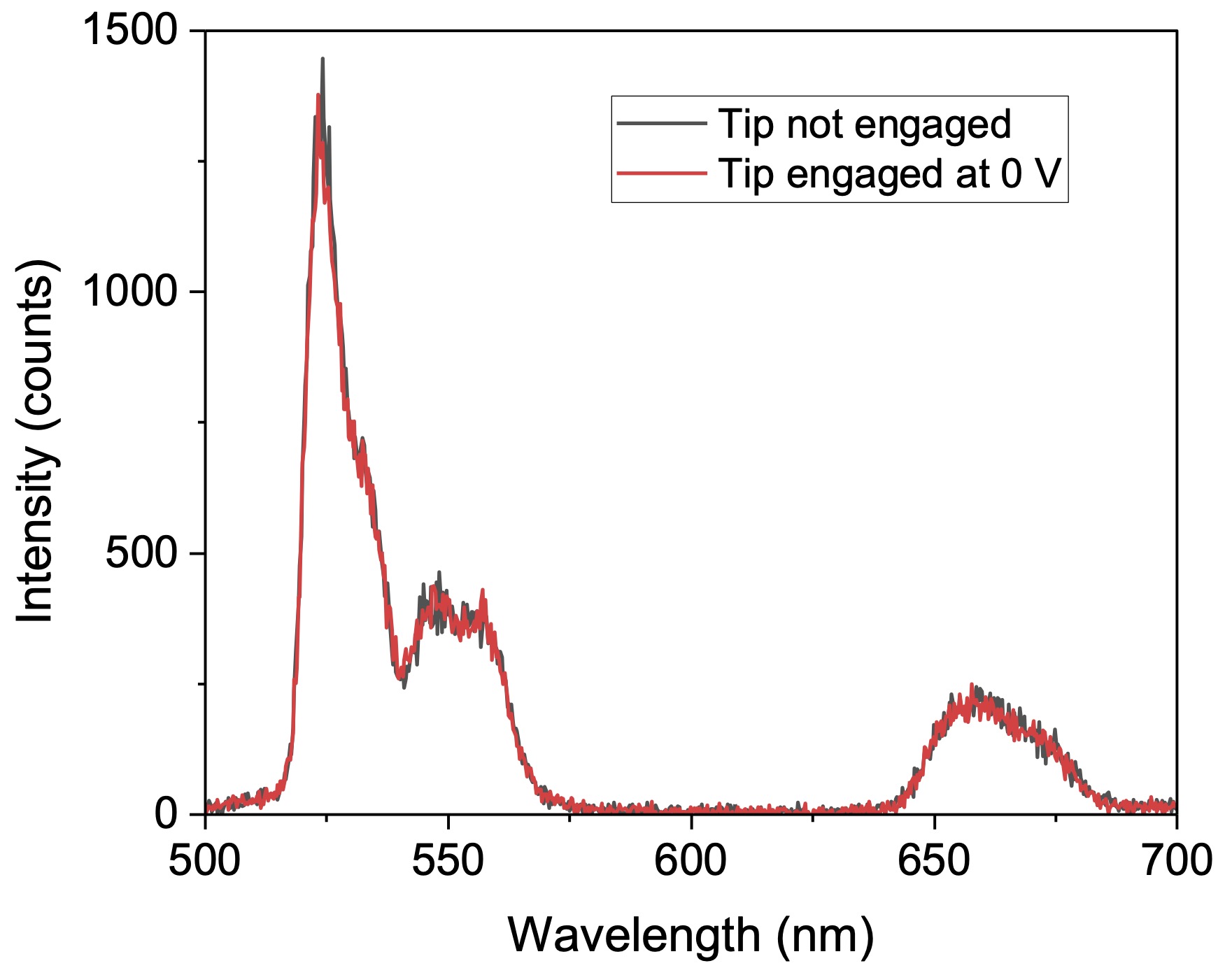}
	\caption{\textbf{Absence of UC spectrum modification on nanoBTO:Yb,Er due to the AFM tip mechanical force applied at the contact}. 
		Spectra recorded at 10~mW excitation laser power and 10~s exposure duration. We do not see any difference between the NC spectrum acquired with the tip not in contact (dark grey) and with the tip in contact with no applied voltage (red).}
	\label{SI_Fig:no_mechanical_effect}
\end{figure}

\begin{figure}[H]
	\centering
	\includegraphics[width=\textwidth]{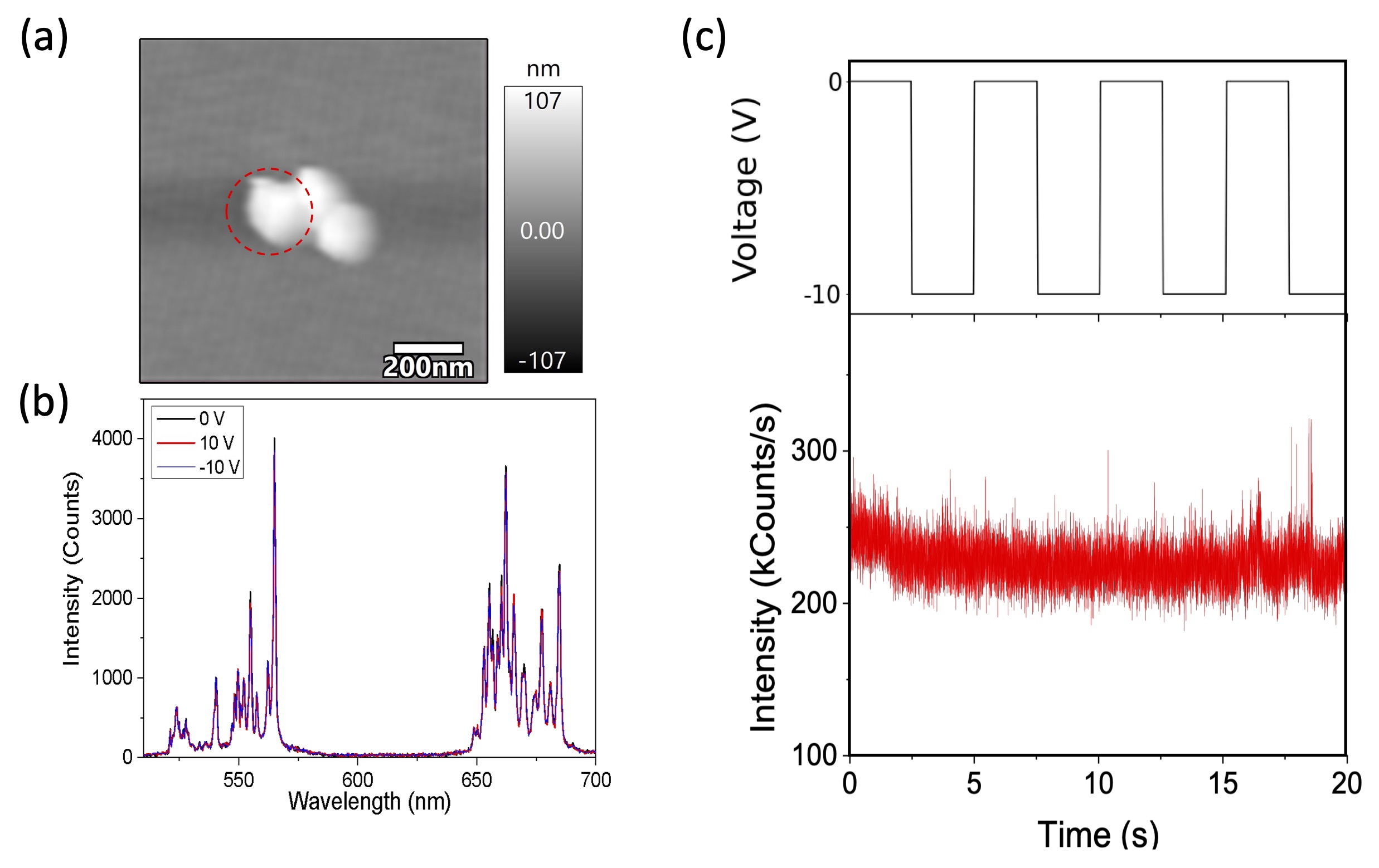}
	\caption{\textbf{Absence of UC spectrum modification under applied voltage in non-ferroelectric Y$_2$O$_3$:Er nanocrystals}. (a) AFM topography image of a small aggregate. The particle surrounded by the dashed circle is the one on top of which the conductive tip is placed to apply voltage. (b) Up-conversion spectra of aggregate shown in (a), at 400~$\mu$W excitation laser (977~nm wavelength) power and with 1~s exposure duration, in various conditions: no applied voltage (black), +10~V applied (red) and -10~V (blue). (c) Total UC intensity time trace during the square variation of applied voltage between 0 and 10~V displayed on top.}
	\label{SI_Fig:nanoY2O3}
\end{figure}

\begin{figure}[H]
	\centering
	\includegraphics[width=\textwidth]{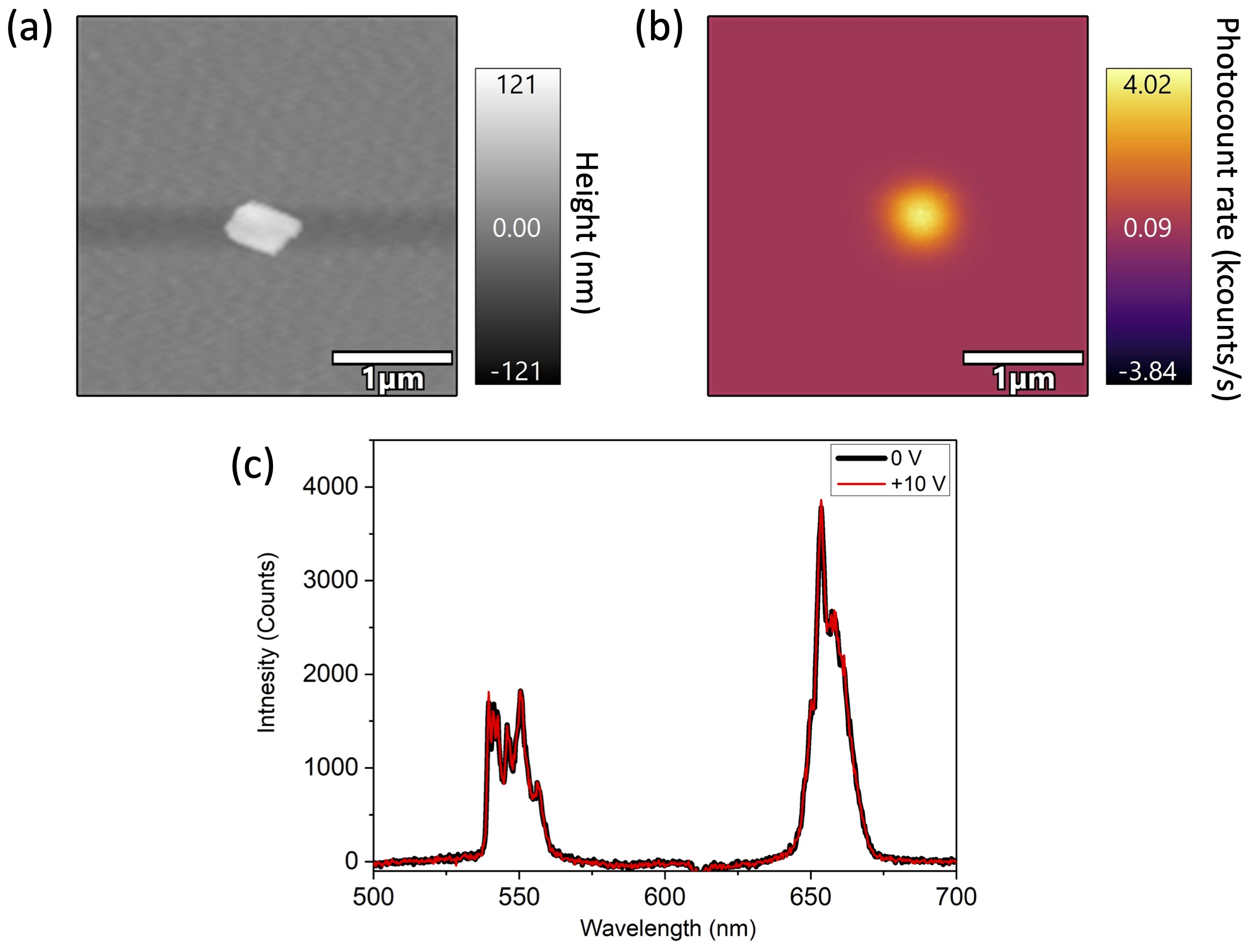}
	\caption{\textbf{Absence of UC spectrum modification under applied voltage in non-ferroelectric NaYF$_4$:Yb,Er nanocrystals}. (a) AFM topography image of a small aggregate of $500\times 65$~nm NaYF$_4$:Yb,Er nanorods. (b) Up-conversion rasterscan of the same aggregate as in (a) at 77~$\mu$W excitation laser (977~nm wavelength) power. (c) UC spectrum at the same excitation laser power as in (b), and 10~s exposure duration, with or without applied 10~V voltage.}
	\label{SI_Fig:nanoNaYF4}
\end{figure}

\begin{figure}[H]
	\centering
	\includegraphics[width=\textwidth]{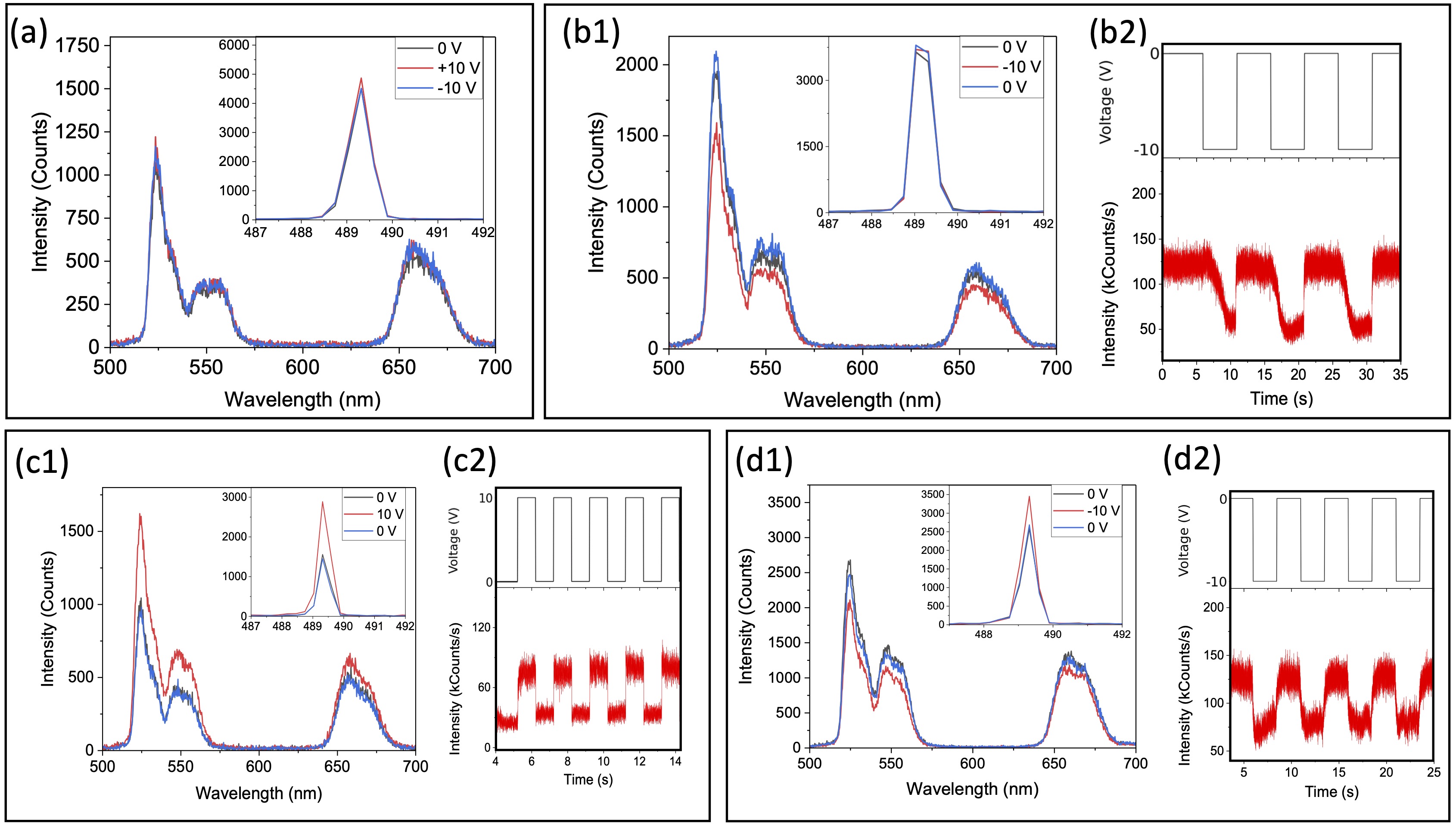}
	\caption{\textbf{Examples of other up-conversion signal variation behaviors with applied electric field, differing from the fast variation with constant SHG.} (a) Example of a nanocrystal with no variation of UC or SHG (inset) intensities whatever the polarity of applied voltage is. This NC has a height of 133~nm. (b) Example of a nanocrystal for which the UC intensity varies (decreases) with a slow response time when -10~V is applied, and without any change in SHG signal. (b1): UC and SHG peak (inset) spectra. (b2): total UC intensity time trace function of applied voltage. (c) Example of a nanocrystal with fast variation of UC intensity and SHG signal. (c1): UC and SHG peak (inset) spectra. (c2): total UC intensity time trace function of applied voltage. (d) Example of a nanocrystal with slow variation of UC intensity and SHG signal. Left (d1): UC and SHG peak (inset) spectra. Right (d2): total UC intensity time trace function of applied voltage.}
	\label{SI_Fig:other_UC_SHG_behaviors}
\end{figure}

\providecommand{\latin}[1]{#1}
\makeatletter
\providecommand{\doi}
{\begingroup\let\do\@makeother\dospecials
	\catcode`\{=1 \catcode`\}=2 \doi@aux}
\providecommand{\doi@aux}[1]{\endgroup\texttt{#1}}
\makeatother
\providecommand*\mcitethebibliography{\thebibliography}
\csname @ifundefined\endcsname{endmcitethebibliography}
{\let\endmcitethebibliography\endthebibliography}{}